\makeatletter
    
    \newcommand{\Rmnum}[1]{\expandafter\@slowromancap\romannumeral #1@}
\makeatother

\documentclass[12pt,journal,draftclsnofoot,a4paper,oneside,onecolumn]{IEEEtran}
\usepackage{amssymb}

\usepackage{amsfonts}
\ifCLASSINFOpdf
  \usepackage[pdftex]{graphicx}
\else
  \usepackage[dvips]{graphicx}
\fi
\usepackage[cmex10]{amsmath}
\usepackage{array}

\usepackage{booktabs}

\usepackage{url}
\begin{document}
\title{Joint Relay and Jammer Selection for Secure Two-Way Relay Networks}


\author{
\IEEEauthorblockN{Jingchao Chen, Rongqing Zhang, Lingyang Song, Zhu
Han\IEEEauthorrefmark{2}, and Bingli~Jiao\\}
\IEEEauthorblockA{\normalsize{School of Electrical Engineering and Computer Science, Peking University, Beijing, China.}\\
                  \IEEEauthorrefmark{2}\normalsize{Electrical and Computer Engineering Department, University of Houston, USA.}}
\thanks{This work is partially supported by US NSF CNS-0910401, CNS-0905556, and CNS-0953377.}}

%\markboth{Journal of \LaTeX\ Class Files,~Vol.~6, No.~1, January~2007}%
%{Shell \MakeLowercase{\textit{et al.}}: Bare Demo of IEEEtran.cls for Journals}



\maketitle

\begin{abstract}
%\boldmath
In this paper, we investigate joint relay and jammer selection in
two-way cooperative networks, consisting of two sources, a number of
intermediate nodes, and one eavesdropper, with the constraints of
physical layer security. Specifically, the proposed algorithms
select two or three intermediate nodes to enhance security against
the malicious eavesdropper. The first selected node operates in the
conventional relay mode and assists the sources to deliver their
data to the corresponding destinations using an amplify-and-forward
protocol. The second and third nodes are used in different
communication phases as jammers in order to create intentional
interference upon the eavesdropper node. Firstly, we find that in a
topology where the intermediate nodes are randomly and sparsely
distributed, the proposed schemes with cooperative jamming
outperform the conventional non-jamming schemes within a certain
transmitted power regime. We also find that, in the scenario in
which the intermediate nodes gather as a close cluster, the jamming
schemes may be less effective than their non-jamming counterparts.
Therefore, we introduce a hybrid scheme to switch between jamming
and non-jamming modes. Simulation results validate our theoretical
analysis and show that the hybrid switching scheme further improves
the secrecy rate.
\end{abstract}


%\begin{IEEEkeywords
%Secure communications,relay selection, two-way.
%\end{IEEEkeywords}

% \fi
\IEEEpeerreviewmaketitle
\newpage

\section{Introduction}

Traditionally security in wireless networks has been mainly focused
on higher layers using cryptographic methods \cite{Silva2008}.
Pioneered by Aaron Wyner's work \cite{Ref1}, which introduced the
wiretap channel and established fundamental results of creating
perfectly secure communications without relying on private keys,
physical-layer-based security has drawn increasing attention
recently. The basic idea of physical layer security is to exploit
the physical characteristics of the wireless channel to provide
secure communications. The security is quantified by the
\emph{secrecy capacity}, which is defined as the maximum rate of
reliable information sent from the source to the intended
destination in the presence of eavesdroppers. Wyner showed that when
the eavesdropper channel is a degraded version of the main channel,
the source and the destination can exchange secure messages at a
non-zero rate. The
following research work \cite{Ref2} studied the
secrecy capacity of the Gaussian wiretap channel, and \cite{Ref3} extended
Wyner's approach to the transmission of confidential messages over
the broadcast channels. Very recently, physical layer security have
been generalized to investigate wireless fading channels
\cite{Ref4}--\cite{Ref7}, and various multiple access scenarios
\cite{Refinsert1}--\cite{Refinsert4}.

Note the fact that if the source-wiretapper channel is stronger than
the source-destination channel, the perfect secrecy rate will be
zero \cite{Ref3}. Some work \cite{Ref8}--\cite{Ref13} has been
proposed to overcome this limitation with the help of relay
cooperation by \emph{cooperative relaying} \cite{Ref8}--\cite{Ref9},
and \emph{cooperative jamming}
\cite{Yener2006Allerton}--\cite{Yener2007France}. For instance, in
\cite{Ref8} and \cite{Ref9}, the authors proposed effective
decode-and-forward (DF) and amplify-and-forward (AF) based
cooperative relaying protocols for physical layer security,
respectively. Cooperative jamming is another approach to improve the
secrecy rate by interfering the eavesdropper with codewords
independent of the source messages. In Yener and Tekin's work
\cite{Yener2006Allerton}, a scheme termed \emph{collaborative
secrecy} was proposed, in which a non-transmitting user was selected
to help increase the secrecy capacity for a transmitting user by
effectively ``jamming'' the eavesdropper. Following similar idea as
\cite{Yener2006Allerton}, they first proposed cooperative jamming in
\cite{Yener2007} and \cite{Yener2007France} in order to increase
achievable rates in the scenarios where general gaussian multiple
access wire-tap channel and two-way wire-tap channel were assumed,
respectively. The authors of \cite{Ekrem2011} and \cite{Ekrem2009}
investigated the effects of user cooperation on the secrecy of broadcast channels by considering a cooperative relay broadcast channel, and showed that user cooperation can increase the achievable secrecy region. The study of communicating through unauthenticated
intermediate relays between a source-destination pair started from
Yenner and He's work \cite{YenerHe2007Asilomar}--\cite{YenerHe2008Asilomar}. The relay
channel with confidential messages was also investigated in
\cite{Ref12}--\cite{Ref13}, where the untrusted relay node acts both
as an eavesdropper and a conventional assistant relay.

Two-way communication is a common scenario in which two nodes
transmit information to each other simultaneously. Recently, the
two-way relay channel \cite{Ref14}--\cite{HeYener2008Asilomar} has
attracted lots of interest from both academic and industrial
communities due to its bandwidth efficiency and potential
application to cellular networks and peer-to-peer networks. In
\cite{Ref14} and \cite{Ref15}, both AF and DF protocols for one-way
relay channels were extended to general full-duplex discrete two-way
relay channel and half-duplex Gaussian two-way relay channel,
respectively. In \cite{Ref16}, network and channel coding were used
in two-way relay channel to increase the sum-rate of two sources.
The work in \cite{Ref17} introduced a two-way memoryless system with
relays in which the signal transmitted by the relay was obtained by
applying an instantaneous relay function to the previously received
signal in
order to optimize the symbol error rate performance. As for the secure communications, in \cite{HeYener2008Asilomar}, Yener and He investigated  the role of feedback in secrecy for two-way networks, and proved that the loss in secrecy rate when ignoring the feedback is very limited in a scenario with half-duplex Gaussian two-way relay channels and an eavesdropper. %In%\cite{ZLCC-2009}, two-way relay channel with linear processing based
%on analogue network coding was analyzed and an optimal relay
%beamforming structure was introduced.

It is well known that, in a cooperative communication network,
proper relay/jammer selection can have a significant impact on the
performance of the whole system. Several relay selection techniques
\cite{jing2009}--\cite{Alam2010} have been explored by far. The
authors in \cite{jing2009} proposed a non-jamming relay selection
scheme for two-way networks with multiple AF relays in an
environment without eavesdroppers, which maximized the worse
received signal-to-noise ratio (SNR) of the two end users. In
\cite{Krikidis2009}, several relay selection techniques were
proposed in one-way cooperative networks with secrecy constraints.
In \cite{Alam2010}, the authors investigated some relay selection
techniques in a two-hop DF cooperative communication system with no
central processing unit to optimally select the relay. Although
cooperative networks have received much attention by far, the
physical layer security issues with secrecy constraints in two-way
schemes have not yet been well investigated.

To this end, in this paper, we propose a scheme that can implement
information exchange in the physical layer against eavesdroppers for
two-way cooperative networks, consisting of two sources, a number of
intermediate nodes, and one eavesdropper, with the constraints for
physical layer security. Unlike \cite{jing2009}, in which the relay
selection is operated in an environment with no security
requirement, our work takes into account the secrecy constraints. In
contrast to \cite{Krikidis2009}, where many relay selections based
on the DF strategy for one-way cooperative wireless networks were
proposed and a safe broadcast phase was assumed, the problem we
consider here involves a non-security broadcast phase, and the
information is transferred bidirectionally.

Specifically, a node is selected from an intermediate node set to
operate at a conventional relay mode, and then uses an AF strategy
in order to assist the sources to deliver data to the corresponding
destination. Meanwhile, another two intermediate nodes that perform
as jammers are selected to transmit artificial interference in order
to degrade the eavesdropper links in the first and second phases of
signal transmissions, respectively. We assume that both destinations
cannot mitigate artificial interference, and thus, the jamming will
also degrade the desired information channels. The principal
question here is how to select the relay and the jamming nodes in
order to increase information security, and meanwhile protect the
source message against eavesdroppers. Several selection algorithms
are proposed, aiming at promoting the assistance to the sources as
well as the interference to the eavesdropper.

The theoretical analysis and simulation results reveal that the
proposed jamming schemes can improve the secrecy rate of the system
by a large scale, but only within a certain transmitted power range.
In some particular scenarios, the proposed schemes become less
efficient than the conventional ones. We then propose a hybrid
scheme with an intelligent switching mechanism between jamming and
non-jamming modes to solve this problem.

The rest of this paper is organized as follows. In Section
\Rmnum{2}, we describe the system model, and formulate the problem
under consideration. Section \Rmnum{3} presents the proposed
selection techniques, and introduces their hybrid implementations.
In Section \Rmnum{4}, we provide both quantitative analysis and
qualitative discussions of different selection schemes in some
typical configurations. Numerical results are shown in Section
\Rmnum{5}, and in Section \Rmnum{6}, we draw the main conclusions.


\section{System Model and Problem Formulation}

\subsection{System Model}

\begin{figure}[!t]
\centering
\includegraphics[width=4.8in]{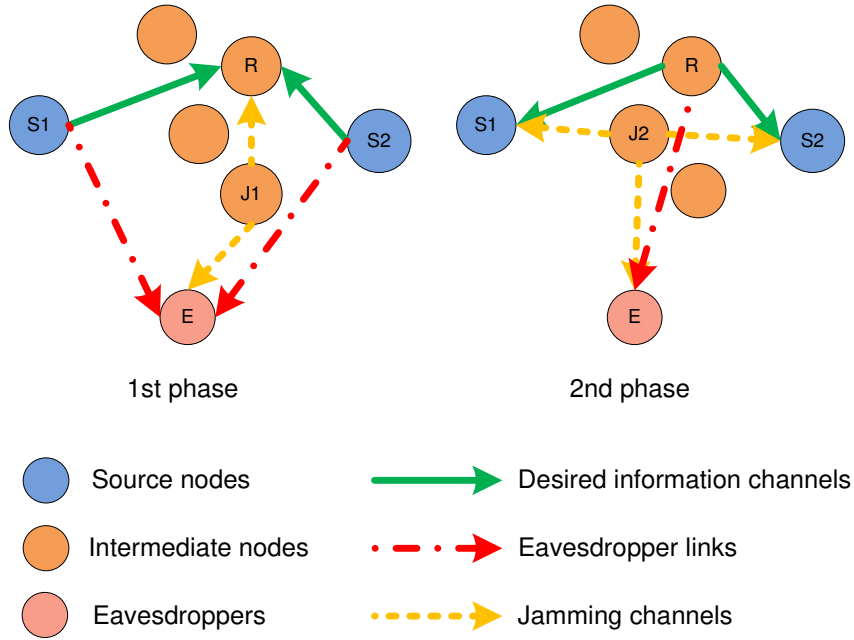}
\caption{System model, where the eavesdropper node is able to
receive signals from both $S_1$ and $S_2$.} \label{fig_1}
\end{figure}

We assume a network configuration consisting of two sources $S_1$
and $S_2$, one eavesdropper $E$, and an intermediate node set
$S_{in}=\left\{{1,2,...,K}\right\}$ with $K$ nodes. In
Fig.~\ref{fig_1} it schematically shows the system model. As the
intermediate nodes cannot transmit and receive simultaneously (half
duplex constraint), the communication process is performed into two
phases. During the broadcasting phase, $S_1$ and $S_2$ transmit
their data to the intermediate nodes. In addition, according to the
security protocol, one node $J_1$ is selected from $S_{in}$ to
operate as a ``jammer'' and transmit intentional interference to
degrade the eavesdropper links in this phase. Since the jamming
signal is unknown at the rest nodes of $S_{in}$, the interference
will also degrade the performance of the relay links, as shown in
Fig.~\ref{fig_1}. In the second phase, according to the security
protocol, an intermediate node $R$ is selected to operate as a
conventional relay and forwards the source messages to the
corresponding destinations. A second jammer $J_2$ is selected from
$S_{in}$, for the same reason as that for $J_1$. Note that the
destinations $S_1$ and $S_2$ are not able to mitigate the artificial
interference from the jamming node, either.

In both two phases, a slow, flat, and block Rayleigh fading
environment is assumed, i.e., the channel remains static for one
coherence interval and changes independently in different coherence
intervals with a variance $\sigma _{i,j}^2  = d_{i,j}^{ - \beta }$,
where $d_{i,j}$ denotes the Euclidean distance between node $i$ and
node $j$, and $\beta$ represents the path-loss exponent. The channel
between node $i$ and node $j$ is denoted as $h_{i,j}$, which is
modeled as a zero-mean, independent, circularly-symmetric complex
Gaussian random variable with variance $ \sigma _{i,j}^2 $.
Furthermore, additive white Gaussian noise (AWGN) with zero mean and
unit variance is assumed. Let $P_S$, $P_R$ and $P_J$ denote the
transmitted power for the source nodes, the relay node and the
jamming nodes, respectively. In order to protect the destinations
from severe artificial interference, the jamming nodes transmit with
a lower power than the relay nodes \cite{Krikidis2009}, and thus
their transmitted power can be defined as $P_J  = P_R /L$, where $ L
\gg 1$ denotes the power ratio of the relay to the jammer.

In the first phase, the two sources send information symbols $s_1$
and $s_2$, respectively, which are mapped to a PSK set. The intermediate node $R$ and eavesdropper $E$
thus receive
\begin{align}
r =\sqrt {P_S } h_{S_1,R} s_1  + \sqrt {P_S } h_{S_2,R} s_2
 + \sqrt {P_J} h_{J_1,R}j_1  + v_R,\\
\label{e1} e_1  = \sqrt {P_S } h_{S_1,E} s_1 + \sqrt {P_S }
h_{S_2,E} s_2 +\sqrt{P_J } h_{J_1,E} j_1  + v_E,
\end{align}
where $v_R$ and $v_E$ denote the noise at $R$ and $E$, respectively.

In the second phase, the node $R$ is selected to amplify its
received signal and forward it to $S_1$ and $S_2$, i.e., $R$
broadcasts
\begin{align}\label{t}
t = \alpha \sqrt {P_R } r,
\end{align} where
$
\alpha =\sqrt {\frac{1}{{1 + |h_{S_1,R} |^2 P_S + |h_{S2,R} |^2 P_S
+ |h_{J_1,R} |^2 P_J }}}$.

 Since the destination $S_i$ knows $s_i$ (for $i=1,2$), it can cancel the self-interference. Therefore,
$S_1$, $S_2$, and the eavesdropper $E$ get
\begin{align}
x_1  =&~ \alpha \sqrt {P_R } \sqrt {{P_S}} {h_{S_2,R}}{h_{R,S_1}}{s_2} + \alpha \sqrt {P_R } \sqrt {{P_J}} {h_{J_1,R}}{h_{R,S_1}}{j_1} \nonumber\\
  &+ \sqrt {{P_J}} {h_{J_2,S_1}}{j_2} + \alpha \sqrt {P_R } {h_{R,S_1}}{v_R} + {w_1}, \\
x_2 =&~ \alpha \sqrt {P_R } \sqrt {{P_S}} {h_{S_1,R}}{h_{R,S_2}}{s_1} + \alpha \sqrt {P_R } \sqrt {{P_J}} {h_{J_1,R}}{h_{R,S_2}}{j_1} \nonumber\\
  &+ \sqrt {{P_J}} {h_{J_2,S_2}}{j_2} + \alpha \sqrt {P_R } {h_{R,S_2}}{v_R} +
  {w_2},\\
\label{e2}
 e_2 =&~ \alpha \sqrt {P_R } \sqrt {P_S } \left( {h_{S_1,R} s_1  +
h_{S_2,R} s_2 } \right)h_{R,E}  + \alpha \sqrt {P_R } h_{R,E}
 v_R \nonumber \\
 &~+ \alpha \sqrt {P_R } \sqrt {P_J} h_{J_1,R} h_{R,E} j_1  + \sqrt {P_J} h_{J_2,E} j_2  +
 w_E,
\end{align}
where $w_1$,$w_2$, and $w_E$ represent the noise terms at $S_1$,
$S_2$, and $E$, respectively. Then, $\Gamma_j$, defined as the
overall signal to interference-plus-noise ratio (SINR) of the
channel $S_i \to S_j$ (for $i,j=1, 2$, $i\neq j $), can be
calculated as
\begin{align} \label{Gamma_j}
\Gamma_j = \frac{{{\gamma _{S_i,S_j}}}}{{{\gamma _{J_1,S_j}} +
{\gamma _{J_2,S_j}} + {\gamma _{R,S_j}} + 1}},
\end{align}
where $\gamma _{m,n}$ represents the instantaneous signal-to-noise
ratio (SNR) for the link $m \to n$:
\begin{align}\label{gamma_SiSj}
 {\gamma _{S_i,S_j}} &= {\alpha ^2}{P_R}{P_S}|{h_{S_i,R}}{|^2}|{h_{R,S_j}}{|^2}, \\
 \label{gamma_J1Sj}
 {\gamma _{J_1,S_j}} &= {\alpha ^2}{P_R}{P_J}|{h_{J_1,R}}{|^2}|{h_{R,S_j}}{|^2}, \\
  \label{gamma_J2Sj}
 {\gamma _{J_2,S_j}} &= {P_J}|{h_{J_2,S_j}}{|^2}, \\
 \label{gamma_RSj}
 {\gamma _{R,S_j}} &= {\alpha ^2}{P_R}|{h_{R,S_j}}{|^2}.
 \end{align}

Strictly speaking, in order to maximize the overall SINR of the
eavesdropping links, the eavesdropper ($E$) can perform whatever
operations as it wishes with the signals received in the previous
two phases. Here in this paper, we take a simple case in which the
eavesdropper applies maximal ratio combining
(MRC)~\cite{TheodoreWireless}, so as to examine the efficiency of
the proposed jamming schemes \footnote{Please note that the
eavesdropper's operation is not limited to maximal ratio combining
(MRC). And the increasing in secrecy rates of the proposed schemes
can still be achieved if the eavesdropper takes other operations,
since the basic forms of the SINRs and thus of the secrecy rates do
not change.}. According to MRC, $E$ combines the received signals by
multiplying $e_1$ in (\ref{e1}) and $e_2$ in (\ref{e2}) with proper
weighting factors $a_1$ and $a_2$, respectively. Without loss of
generality, consider the scenario in which $E$ intends to optimize
the SINR of eavesdropper link $S_i \to E$, for $i=1, 2$, the
combined eavesdropping signal can be written as
\begin{align}
 {e^i} = a_1^i{e_1} +a_2^i{e_2},
\end{align}
where
\begin{align}
a_1^i&\buildrel \Delta \over = \frac{{\sqrt
{{P_S}} h_{S_i,E}^H}}{{\sigma _{N_{e_1},S_j}^2}},\\
a_2^i &\buildrel \Delta \over = \frac{{\alpha \sqrt {{P_S}}
h_{S_i,R}^H h_{R,E}^H}}{{\sigma _{N_{e_2},S_j}^2}},
\end{align}
with $i,j=1,2,i\neq j $, and $(\cdot)^H$ is the conjugate transpose.
$\sigma _{N_{e_1},S_j}^2$ and $\sigma _{N_{e_2},S_j}^2$ represent
the total interference and noise power terms in $e_1$ and $e_2$, respectively:
\begin{align}
\sigma _{N_{e_1},S_j}^2 &= {\gamma _{S_j,E}} + {\gamma _{J_1,E}} + 1, \\
\sigma _{N_{e_2},S_j}^2 &= {\gamma _{S_j,R,E}} + {\gamma _{J_1,R,E}}
+ {\gamma _{J_2,E}} + {\gamma _{R,E}} + 1,
\end{align}
where
\begin{align}
 &{\gamma _{S_j,E}} = {P_S}|{h_{S_j,E}}{|^2}, \\
 &{\gamma _{J_j,E}} = {P_J}|{h_{J_j,E}}{|^2}, \\
 \label{gamma_SjRE}
 &{\gamma _{S_j,R,E}} = {\alpha ^2}{P_R}{P_S}|{h_{S_j,R}}{|^2}|{h_{R,E}}{|^2}, \\
 \label{gamma_J1RE}
 &{\gamma _{J_1,R,E}} = {\alpha ^2}{P_R}{P_J}|{h_{J_1,R}}{|^2}|{h_{R,E}}{|^2},\\
 &{\gamma _{R,E}} = {\alpha ^2}{P_R}|{h_{R,E}}{|^2}.
\end{align}

In order to calculate the SINR of link $S_i \to E$, we assume two
different channel knowledge sets:
\begin{enumerate}
\item $\psi _0$ that denotes a global instantaneous knowledge for all the links,
\item $\psi _1$ that denotes an average channel knowledge for the
eavesdropper links.
\end{enumerate}

With the assumption of $\psi_0$, we can get the instantaneous SNR of
any channel $i \to j$ in the system. Thus, the SINR of link $S_i \to
E$ can be calculated as
\begin{align} \label{Gamma_Ei}
\Gamma _{E_i }&= \frac{{{P_S}|{h_{S_i,E}}{|^2}}}{{\sigma
_{N_{e_1},S_j}^2}} + \frac{{{\alpha
^2}{P_R}{P_S}}|{h_{S_i,R}}{|^2}|{h_{R,E}}{|^2}}{\sigma
_{N_{e_2},S_j}^2}\nonumber \\
&= \frac{{{\gamma _{S_i,E}}}}{{{\gamma _{S_j,E}} + {\gamma _{J_1,E}}
+ 1}} + \frac{{{\gamma _{S_i,R,E}}}}{{{\gamma _{S_j,R,E}} + {\gamma
_{J_1,R,E}} + {\gamma _{J_2,E}} + {\gamma _{R,E}} + 1}},
\nonumber \\
&\mbox{s.t.}~\psi_0.
\end{align}

In an environment where the instantaneous channel knowledge set
$\psi_0$ is not available, we can use the expectation of SNRs for
the eavesdropper links $\mathbb{E}\left[\gamma _{S_i,E}\right]$,
which is provided by the average channel knowledge $\psi_1$ , to get
the SINRs:

\begin{align} \label{A_Gamma_Ei}
\Gamma'_{E_i} &= \frac{{\mathbb{E}\left[ {{\gamma _{S_i,E}}}
\right]}}{{\mathbb{E}\left[ {{\gamma _{S_j,E}}} \right] +\mathbb{E}
\left[ {{\gamma_{J_1,E}}} \right] + 1}} +\frac{{\mathbb{E}\left[
{{\gamma _{S_j,R,E}}} \right]}}{{\mathbb{E}\left[ {{\gamma
_{S_j,R,E}}} \right] + \mathbb{E}\left[ {{\gamma _{J_1,R,E}}}
\right] + \mathbb{E}\left[ {{\gamma _{J_2,E}}} \right] +
\mathbb{E}\left[ {{\gamma _{R,E}}} \right] + 1}},
\nonumber \\
&\mbox{s.t.}~\psi_1,
\end{align}
where $ \mathbb{E}\left[ \cdot \right]$ stands for the expectation
operator.

\subsection{Problem Formulation}

The instantaneous secrecy rate for the node set $S_{in}$ for source
$S_i$ can be expressed \cite{Ref11}
\begin{align}
{R_{S_i}} \left( {R,{J_1},{J_2}} \right) = \left[ \frac{1}{2}{{\log
}_2}\left( {1 + \Gamma_i} \right) - \frac{1}{2}{{\log }_2}\left( {1
+ \Gamma_{E_j}} \right) \right]^ +,
\end{align}
where $i=1,2$, $j=1,2$, $i\neq j$, and $ \left[ x \right]^ +
\buildrel \Delta \over = \max \left\{ {0,x} \right\}$.

The overall secrecy performance of the system  is characterized by
the ergodic secrecy capacity that is the expectation of the sum of
the two sources' secrecy rates, $\mathbb{E}\left[R_S \left( {R,J_1
,J_2 } \right)\right]$, where
\begin{align} \label{Cs}
R_S \left( {R,J_1 ,J_2 } \right)= R_{S_1} \left( {R,J_1 ,J_2 }
\right) + R_{S_2} \left( {R,J_1 ,J_2 } \right).
\end{align}

Our objective is to select appropriate nodes $R$, $J_1$, and $J_2$
in order to maximize the instantaneous secrecy rate subject to
different types of channel feedback. The optimization problem can be
formulated as
\begin{align} \label{selecriterion}
\left( {R^* ,J_1^* ,J_2^* } \right) &= \mathop {\arg \max
}\limits_{\scriptstyle R, J_1, J_2  \in S_{in}  \hfill \atop
  \scriptstyle R \ne J_1 ,J_2  \hfill} R_S \left( {R,J_1 ,J_2 }
\right),
\nonumber \\
&\mbox{s.t.}~\psi_u,
\end{align}
where $u=0,1$; $R^*$, $J_1 ^*$ and $J_2 ^*$ denote the selected
relay and jamming nodes, respectively. Note that here the selected
jammers $J_1^*$ and $J_2^*$ in the two phases may be the same node,
which is determined by the instantaneous secrecy rate.

\subsection{Selection without Jamming}

In a conventional cooperative network, the relay scheme does not
have the help from jamming nodes. We derive the following solutions
under this scenario.

\subsubsection{Conventional Selection (CS)}

The conventional selection does not take the eavesdropper channels
into account, and the relay node is selected according to the
instantaneous SNR of the links between node $S_1$ and node $S_2$ only. Therefore, the SINR given in (\ref{Gamma_j}) becomes
\begin{align} \label{Gamma_j_cs}
\Gamma_j^{CS} = \frac{{{\gamma _{S_i,S_j}}}}{{ {\gamma _{R,S_j}} + 1}},
\end{align}
where $\Gamma_j^{CS}$ represents the SINR of the
channel $S_i \to S_j$ (for $i,j=1, 2$, $i\neq j $) without considering the eavesdropper.

Hence, the conventional
selection algorithm can be expressed as
\begin{align} \label{CS}
 R^*  &= \arg \mathop {\max }\limits_{R \in S_{in} } \left\{ {R_{S_1} \left( R \right) + R_{S_2} \left( R \right)} \right\}\nonumber \\
 &= \arg \mathop {\max }\limits_{R \in S_{in} } \left\{\frac{1}{2} {\log_2\left(1+\Gamma_1 ^{CS}\right)  + \frac{1}{2}\log_2\left(1+ \Gamma_2 ^{CS}\right) } \right\} \nonumber \\
  &= \arg \mathop {\max }\limits_{R \in S_{in} } \left\{ {\left(1+\frac{{\gamma _{S_1,S_2} }}{{\gamma _{R,S_2}  + 1}}\right) \cdot \left(1+ \frac{{\gamma _{S_2,S_1} }}{{\gamma _{R,S_1}  + 1}} \right)}
  \right\},
\end{align}
with $\gamma _{S_i,S_j}$ and $\gamma _{R,S_j}$ for $\left( {i,j =
0,1} \right)$ given by (\ref{gamma_SiSj}) and (\ref{gamma_RSj}),
respectively. Since (\ref{CS}) shows that this selection does not consider the eavesdropping links, the CS algorithm may not able to support systems with the secrecy constraints even though it
may effective in non-eavesdropper environments.

\subsubsection{Optimal Selection (OS)}

This solution takes the eavesdropper into account and selects the
relay node based on $\psi _0$, which provides the instantaneous
channel knowledge for all the links. Then, the SINR of link $S_i \to E$ in (\ref{Gamma_Ei}) can be rewritten as
\begin{align}
\label{Gamma_OS_Ei}
 \Gamma^{OS}_{E_i}  &=\frac{{\gamma _{S_i,E}
}}{{\gamma _{S_j,E}+1 }} + \frac{{\gamma _{S_i,R,E} }}{{\gamma
_{S_j,R,E}  + \gamma _{R,E}+1 }}.
\end{align}

The optimal selection is given
as:
\begin{align} \label{OS}
R^*  &= \arg \mathop {\max }\limits_{R \in S_{in} } \left\{ {R_{S_1} \left( R \right) + R_{S_2} \left( R \right)} \right\}\nonumber \\
      &= \arg \mathop {\max }\limits_{R \in S_{in} } \left\{\frac{1}{2} {\log_2\left(1+\Gamma_1 ^{OS}\right) -\frac{1}{2}{{\log }_2}\left( {1
+ \Gamma^{OS}_{E_2}} \right) + \frac{1}{2}\log_2\left(1+ \Gamma_2
^{OS}\right)-\frac{1}{2}{{\log }_2}\left( {1
+ \Gamma^{OS}_{E_1}} \right) } \right\} \nonumber \\
     &=\arg
\mathop {\max }\limits_{R \in S_{in} } \left\{
{\frac{{1+\Gamma^{OS}_1
 }}{{1+\Gamma^{OS}_{E_2}  }} \cdot \frac{{1+\Gamma^{OS}_2
}}{{1+\Gamma^{OS}_{E_1}  }}} \right\},
\end{align}
where
\begin{align}
\label{Gamma_OS_i}
\Gamma^{OS}_i  &=\Gamma^{CS}_i=\frac{{\gamma _{S_j,S_i} }}{{\gamma _{R,S_i}+1 }}.
\end{align}

\subsubsection{Suboptimal Selection (SS)}

The suboptimal selection implements the relay selection based on the
knowledge set $\psi _1$, which gives the average estimate of the
eavesdropper links. Therefore, it avoids the difficulty of getting
instantaneous estimate of the channel feedbacks. Similar to the OS algorithm in (\ref{OS}), the suboptimal
selection can be written as
\begin{align}\label{SS}
R^* =  \arg \mathop {\max }\limits_{R \in S_{in} } \left\{
{\frac{{1+\Gamma^{SS}_1  }}{{1+\Gamma ^{SS}_{E_2} }} \cdot
\frac{{1+\Gamma^{SS}_2  }}{{1+\Gamma ^{SS}_{E_1} }}} \right\},
\end{align}
where
\begin{align} \label{Gamma_i_SS}
\Gamma^{SS}_i   &= \Gamma^{OS}_i   =\frac{{\gamma _{S_j,S_i}
}}{{\gamma _{R,S_i}+1 }},\\
\label{Gamma_Ei_SS} \Gamma^{SS}_{E_i} &=\frac{{\mathbb{E}\left[
{\gamma _{S_i,E} } \right]}}{{\mathbb{E}\left[ {\gamma _{S_j,E} }
\right]+1}} + \frac{{\mathbb{E}\left[ {\gamma _{S_i,R,E} }
\right]}}{{\mathbb{E}\left[ {\gamma _{S_j,R,E} } \right] +
\mathbb{E}\left[ {\gamma _{R,E} } \right]+1}}.
\end{align}
Note that in comparison of the OS in (\ref{OS}), the only difference of the SS algorithm in (\ref{SS}) is that it requires the average channel state information, $\psi _1$, which would be more useful in practice.

\section{Selections with Jamming in Two-way Relay Systems}
In this section, we present several node selection techniques based
on the optimization problem given by (\ref{selecriterion}) in the
two-way systems. Unlike \cite{Krikidis2009}, where the selection
techniques only concern about the secrecy performance in the second
phase of transmission, here, our work takes into account both the
two phases in order to select a set of relay and jammers that can
maximize the overall expectation of secrecy rate.

\subsection{Optimal Selection with Maximum Sum Instantaneous Secrecy Rate (OS-MSISR) }

The optimal selection with maximum sum instantaneous secrecy rate
assumes a knowledge set $\psi_0$ and ensures a maximization of the
sum of instantaneous secrecy rates of node $S_1$ and node $S_2$
given in (\ref{Cs}), which gives credit to
\begin{align} \label{OS-MSISR_selection}
\left( {{R^*},{J_1}^*,{J_2}^*} \right)
 &=\mathop {\arg\max }\limits_{\scriptstyle R, {J_1}, {J_2} \in {S_{in}} \hfill
\atop
  \scriptstyle R \neq  {J_1},{J_2} \hfill} \left\{ {R_S}(R,{J_1},{J_2}) \right\}\nonumber\\
&= \mathop {\arg \max }\limits_{\scriptstyle R, {J_1}, {J_2} \in
{S_{in}} \hfill \atop
  \scriptstyle R \neq  {J_1},{J_2} \hfill} \left\{ {\frac{{1 + \Gamma_2}}{{1 + \Gamma_{E_1}}} \cdot \frac{{1 + \Gamma_1}}{{1 + \Gamma_{E_2}}}}
  \right\},
\end{align}
where $\Gamma_i$ and $\Gamma_{E_i}$ are given by (\ref{Gamma_j}) and
(\ref{Gamma_Ei}), respectively.

The approach in (\ref{OS-MSISR_selection}) reflects the basic idea of using both
cooperative relaying and cooperative jamming in order to promote the
system's secrecy performance. Specifically, the OS-MSISR scheme here
tends to select a set of relay and jammers that maximizes
$\Gamma_i$, which means promoting the assistance to the sources.
Meanwhile this relay and jammer set tends to minimize
$\Gamma_{E_i}$, which is equivalent to enhance the interference to
the eavesdropper.

Although the OS-MSISR scheme seems to be a straightforward
application for cooperative relaying and cooperative jamming, the
actual selection procedure usually involves trade-offs. For
instance, according to (\ref{Gamma_j}) and (\ref{gamma_J1Sj}), we
should select the relay and jammer set that
 minimizes $|h_{J_1,R}|$  in order to make
$\Gamma_i$ as high as possible. Considering (\ref{gamma_SjRE}),
(\ref{gamma_J1RE}) and (\ref{Gamma_Ei}), however, the lower
$|h_{J_1,R}|$ is, the higher $\Gamma_{E_i}$ is, which is undesirable.
Thus, we have to make a trade-off between raising $\Gamma_i$ and
inhibiting $\Gamma_{E_i}$ in order to optimize the right part of
(\ref{OS-MSISR_selection}).

%For high SINRs, $\Gamma_{j,E_i} \gg 1$, (\ref{OS-MSISR_selection})
%can be simplified as
%\begin{align} \label{OS-MSISR_selection_approx}
% \left( {{R^*}, {J_1}^*, {J_2}^*} \right)
%\approx \mathop {\arg \max }\limits_{\scriptstyle R, {J_1}, {J_2} \in
%{S_{in}} \hfill \atop
%  \scriptstyle R \neq {J_1}, {J_2} \hfill} \left\{ \frac{\Gamma_2}{\Gamma_{E_1}} \cdot \frac{\Gamma_1}{\Gamma_{E_2}}
%  \right\}.
%\end{align}

\subsection{Optimal Selection with Max-Min Instantaneous Secrecy Rate (OS-MMISR)}

It is obvious that the OS-MSISR in (\ref{OS-MSISR_selection}) is
complicated, in this subsection we propose a reduced-complexity
algorithm. It is common that the sum secrecy rate of two sources,
i.e. $R_{S_1} \left( {R,J_1 ,J_2 }\right) + R_{S_2} \left( {R,J_1
,J_2 }\right)$, may be driven down to a low level by the user with
the lower secrecy rate. As a result, for low complexity, the
intermediate nodes, which maximize the minimum secrecy rate of two
users, can be selected to achieve the near-optimal performance. In
addition, in some scenarios, the considered secrecy performance does
not only take into account the total secrecy rate of all the source
nodes, but also the individual secrecy rate of each node. If one
source node has low secrecy rate, the whole system is regarded as
secrecy inefficient. Furthermore, assuring each individual source
node a high secrecy rate is another perspective of increasing the
whole system's secrecy performance.

The OS-MMISR selection maximizes the
worse instantaneous secrecy rate of the two source nodes with the
assumption of knowledge set $\psi_0$, and we can get
\begin{align} \label{OS-MMISR_selection}
\left( {{R^*},{J_1}^*,{J_2}^*} \right)
 &=\mathop {\arg \max \min }\limits_{\scriptstyle R,{J_1},{J_2} \in {S_{in}} \hfill \atop
  \scriptstyle R \neq {J_1},{J_2} \hfill} \left\{ {R_{S_1}(R,{J_1},{J_2}),~R_{S_2}(R,{J_1},{J_2})}
  \right\} \nonumber \\
 &=\mathop {\arg \max \min }\limits_{\scriptstyle R,{J_1},{J_2} \in {S_{in}} \hfill \atop
  \scriptstyle R \neq {J_1},{J_2} \hfill} \left\{ \frac{{1 + \Gamma_2}}{1 + {\Gamma_{E_1}}},~\frac{{1 + \Gamma_1}}{{1 + {\Gamma_{E_2}}}}
  \right\},
 \end{align}
where $\Gamma_i$ and $\Gamma_{E_i}$ are given by (\ref{Gamma_j}) and
(\ref{Gamma_Ei}), respectively.

\subsection{Optimal Switching (OSW)}

The original idea of using jamming nodes is to introduce
interference on the eavesdropper links. However, there are two
side-effects of using jamming. Firstly, the jamming node in the
second phase $J_2$ poses undesired interference directly onto the
destinations. Secondly, it degrades the links between the relay node
$R$ and the destinations. Given the assumption that the destinations
cannot mitigate this artificial interference, continuous jamming in
both phases is not always beneficial for the whole system. In some
specific situation (e.g., $J_2$ is close to one destination), the
continuous jamming may decrease secrecy rate seriously, and act as a
bottleneck for the system. In order to overcome this problem, we
introduce the idea of intelligent switching between the OS-MSISR and
the OS scheme in order to reduce the impact of ``negative
interference''. The threshold for the involvement of the jammer
nodes is
\begin{align} \label{OSW_criterion}
R_{S_1} \left( {R,J_1 ,J_2 } \right) + R_{S_2} \left( {R,J_1 ,J_2 }
\right) > R_{S_1}^{OS} \left( R \right) + R_{S_2}^{OS} \left( R
\right),
\end{align}
where
\begin{align}
 R^{OS}_{S_i} \left( R \right) = \left[
{\frac{1}{2}\log _2 \left( {\frac{{1 + \Gamma^{OS}_i  }}{{1 +
\Gamma^{OS}_{E_j} }}} \right)} \right]^ +.
\end{align}

Thus, (\ref{OSW_criterion}) can be further written as
\begin{align} \label{OSW_criterion_approx}
\frac{{1+\Gamma_1 }}{{1+\Gamma_{E_2} }} \cdot \frac{{1+\Gamma_2
}}{{1+\Gamma_{E_1} }}
> \frac{{1+\Gamma^{OS}_1 }}{{1+\Gamma^{OS}_{E_2} }} \cdot \frac{{1+\Gamma^{OS}_2
}}{{1+\Gamma^{OS}_{E_1} }},
\end{align}
where $\Gamma_i$, $\Gamma_{E_i}$, $\Gamma^{OS}_i$ and
$\Gamma^{OS}_{E_i}$ are given by (\ref{Gamma_j}) and
(\ref{Gamma_Ei}), (\ref{Gamma_OS_i}) and (\ref{Gamma_OS_Ei}),
respectively.

For each time slot, if (\ref{OSW_criterion_approx}) is met, the
OS-MSISR scheme provides higher instantaneous secrecy rate than OS
does and is preferred. Otherwise the OS scheme is more efficient in
promoting the system's secrecy performance, which should be employed. Because of the
uncertainty of the channel coefficient $h_{i,j}$ for each channel $i
\to j$, the OSW should outperform either the continuous jamming
scheme or the non-jamming one.

\subsection{Suboptimal Selection with Maximum Sum Instantaneous Secrecy Rate (SS-MSISR)}

With the assumption of $\psi_0$, we can get some optimal selection
metrics. However, its practical interest and potential implements
are only limited to some special (e.g. military) applications, where
the instantaneous quality of the eavesdropper links can be measured by
some specific protocols. In practice, only an average knowledge of
these links $\psi_1$ would be available from long term eavesdropper
supervision. The selection metrics is modified as

\begin{align} \label{SS-MSISR_selection}
\left( {{R^*},{J_1}^*,{J_2}^*} \right)
 = \mathop {\arg \max }\limits_{\scriptstyle R, {J_1}, {J_2} \in
{S_{in}} \hfill \atop
  \scriptstyle R \neq  {J_1},{J_2} \hfill} \left\{ {\frac{{1 + \Gamma_2}}{{1 + \Gamma'_{E_1}}} \cdot \frac{{1 + \Gamma_1}}{{1 + \Gamma'_{E_2}}}}
  \right\},
\end{align}
where $\Gamma_i$ and $\Gamma'_{E_i}$ are given by (\ref{Gamma_j})
and (\ref{A_Gamma_Ei}), respectively.

From (\ref{SS-MSISR_selection}), we can predict that for a scenario in which the intermediate nodes are sparsely
distributed across the considered area, the SS-MSISR scheme can
provide similar relay and jammer selection with the OS-MSISR scheme.
This is because a slightly difference between $\mathbb{E}\left[
\gamma_{i, E}\right]$ provided by $\psi_1$ and $\gamma_{i, E}$
provided by $\psi_0$ would not be enough for the scheme to select another
far-away intermediate node. Thus, under this condition, the average
eavesdropper channel knowledge set $\psi_1$ may contain sufficient
channel information as well for a quasi-optimal selection.

\subsection{Suboptimal Selection with Max-Min Instantaneous Secrecy Rate (SS-MMISR)}

This scheme refers to the practical application of the above
selection with maximum worse instantaneous secrecy rate in (\ref{OS-MMISR_selection}). The basic
idea of considering $\psi_1$ as the average behavior of eavesdropper
links is the same as SS-MSISR, but aimed at looking for the maximum
worse instantaneous secrecy rate, which is written as
\begin{align} \label{SS-MMISR_selection}
\left( {{R^*},{J_1}^*,{J_2}^*} \right)
 &=\mathop {\arg \max \min }\limits_{\scriptstyle R,{J_1},{J_2} \in {S_{in}} \hfill \atop
  \scriptstyle R \neq {J_1},{J_2} \hfill} \left\{ {R_{S_1}(R,{J_1},{J_2}),~R_{S_2}(R,{J_1},{J_2})}
  \right\} \nonumber \\
 &=\mathop {\arg \max \min }\limits_{\scriptstyle R,{J_1},{J_2} \in {S_{in}} \hfill \atop
  \scriptstyle R \neq {J_1},{J_2} \hfill} \left\{ \frac{{1 + \Gamma_2}}{1 + {\Gamma'_{E_1}}},~\frac{{1 + \Gamma_1}}{{1 + {\Gamma'_{E_2}}}}
  \right\},
 \end{align}
where $\Gamma_i$ and $\Gamma'_{E_i}$ are given by (\ref{Gamma_j})
and (\ref{A_Gamma_Ei}), respectively.

\subsection{Suboptimal Switching (SSW)}

Given the fact that jamming is not always a positive process for the
performance of the system, the suboptimal switching refers to the
practical application of the intelligent switching between the
SS-MSISR and the SS schemes. The basic idea is the same as OSW, but
the switching criterion uses the available knowledge set $\psi_1$.
More specifically, the required condition for switching from
SS-MSISR to SS mode is
\begin{align}
\frac{{1+\Gamma_1 }}{{1+\Gamma'_{E_2} }} \cdot \frac{{1+\Gamma_2
}}{{1+\Gamma'_{E_1} }}
> \frac{{1+\Gamma^{SS}_1 }}{{1+\Gamma^{SS}_{E_2} }} \cdot \frac{{1+\Gamma^{SS}_2
}}{{1+\Gamma^{SS}_{E_1} }},
\end{align}
where $\Gamma_i$, $\Gamma'_{E_i}$, $\Gamma^{SS}_i$ and
$\Gamma^{SS}_{E_i}$ are given by (\ref{Gamma_j}) and
(\ref{A_Gamma_Ei}), (\ref{Gamma_i_SS}) and (\ref{Gamma_Ei_SS}),
respectively.

\subsection{Optimal Selection with ``Known'' Jamming (OSKJ)}

The previous selection techniques are proposed based on the
assumption that the jamming signal is unknown at both the two
destinations. This assumption avoids the initialization period in
which the jamming sequence is defined, and thus, it reduces the risk
of giving out the artificial interference to the eavesdropper. For
comparison reasons, here we propose a ``control'' scheme, in which
the jamming signal can be decoded at the destinations $S_1$ and
$S_2$, but not at the eavesdropper $E$. In this case, the SINR of
the link from $S_i$ (for $i=1,2$) to $E$ remains the same as
$\Gamma_{E_i}$ given by (\ref{Gamma_Ei}). The SINR of the link from
$S_i$ to $S_j$ (for $i,j=1,2, i\neq j$) is modified as follows:
\begin{align}
\Gamma_i  = \frac{{\gamma _{Sj,Si} }}{{\gamma _{R,Si}  + 1}}.
\end{align}

The OSKJ scheme is taken into consideration in the numerical results
section as a reference. This, however, is not the ``ideal'' jamming scheme
since the artificial interference from the jammers only degrades the
eavesdropper links. As we have discovered and will discuss
in Section \Rmnum{5}, in some particular scenarios, the OSKJ scheme
is outperformed by the OSW and SSW schemes presented above, for
the jamming has changed the value of $\alpha$ given in~(\ref{t}).

%The proposed and conventional selection schemes are summarized in
%TABLE \Rmnum{1}.
%
%\begin{table}[htbp]
%\caption{Catalog of different selection schemes} \centering
%\begin{tabular}{c|c|c|c|c|c|c|c}
%    \toprule
%    Selection techniques  & OS-MSISR  & OS-MMISR  & SS-MSISR  & SS-MMISR  & OSW  & SSW  & OSKJ  \\
%    \midrule
%    Selection criteria  & Eq.(37)  & Eq.(40)  & Eq.(37)  & Eq.(40)  & \multicolumn{3}{c}{Eq.(37)} \\
%    \midrule
%    $SINR_i$  & \multicolumn{4}{c}{Eq.(8) } &\multicolumn{2}{c}{Eq.(8)/Eq.(28) } & Eq.(46) \\
%    \midrule
%    $SINR_{Ei}$ & \multicolumn{2}{c}{Eq.(23)} & \multicolumn{2}{c}{Eq.(44)}& Eq.(23)/Eq.(29) & Eq.(44)/Eq.(35) & Eq.(23)  \\
%    \midrule
%    $R_{Si}$ &\multicolumn{7}{c}{Eq.(24)}  \\
%    \bottomrule
%\end{tabular}
%\end{table}

\section{Performance Analysis}

In this section, we firstly do some quantitative analysis on the
asymptotic performance of both the proposed jamming and non-jamming
schemes in high transmitted power range. Then, we provide a
qualitative discussion of the secrecy performance of different
selection schemes in some typical scenarios based on the system
model in Section~\Rmnum{2}.

\subsection{Asymptotic Performance for Selections without
Jamming }

Without loss of generality, we take the OS scheme for example. With
high transmitted power $P_S$, we can get
\begin{align}
\label{Gamma_i_OS_approx}
 \Gamma_i^{OS}  &\to P_S |h_{S_i,R} |^2,  \\
\label{Gamma_Ei_OS_approx}
 \Gamma_{E_i}^{OS}  &\to \frac{{|h_{S_i,E} |^2 }}{{|h_{S_j,E} |^2 }} + \frac{{|h_{S_i,R} |^2 }}{{|h_{S_j,R} |^2
 }}.
 \end{align}

We can see that $\Gamma_i ^{OS}$ grows rapidly as $P_S$ increases,
while $\Gamma_{E_i} ^{OS}$ converges to a value that depends only on
the relative distances between the sources, the eavesdropper and the
relay. Therefore, the ergodic secrecy capacity $\mathbb{E}\left[
{R_S}\right]$ also increases rapidly with the transmitted power
$P_S$. Based on (\ref{Gamma_i_OS_approx}) and
(\ref{Gamma_Ei_OS_approx}), the slope of the curve of
$\mathbb{E}\left[ {R_{S} }\right]$ versus $P_S$ (measured by dB) can
be approximately calculated as
\begin{align}\label{slope_approx}
 &\frac{{\partial \mathbb{E}\left[ {R_{S_1}  + R_{S_2} } \right]}}{{\partial P_S }} \nonumber\\
 =&~ \frac{{\partial \mathbb{E}\left[ {\frac{1}{2}\log _2 \frac{{10^{P_S /10} |h_{S1,R} |^2  \cdot 10^{P_S /10} |h_{S2,R} |^2 }}{{\Gamma_{E1}^{OS}  \cdot \Gamma^{OS}_{E2} }}} \right]}}{{\partial P_S }} \nonumber\\
  =&~ \frac{{\partial \mathbb{E}\left[ {\frac{1}{2}\log _2 10^{2P_S /10} } \right]}}{{\partial P_S }} + \frac{{\partial \mathbb{E}\left[ {\frac{1}{2}\log _2 \frac{{|h_{S1,R} |^2 |h_{S2,R} |^2 }}{{\Gamma^{OS}_{E1}  \cdot \Gamma^{OS}_{E2} }}} \right]}}{{\partial P_S }} \nonumber\\
  =&~ \frac{{\partial \left( {\frac{{P_S }}{{10}}\log _2 10} \right)}}{{\partial P_S }} \nonumber\\
  =&~ \frac{1}{{10}}\log _2 10 \nonumber\\
  \approx&~ 0.3322
 \end{align}

For the other non-jamming schemes (i.e. CS, SS),  we note that they
share the same asymptotic performance as the OS scheme with a linear
increment of slope about 0.3322 as the transmitted power $P_S$
increases.

\subsection{Asymptotic Analysis for Selections with Continuous Jamming}

We use the same method as in the previous analysis for the
non-jamming selections to analyze the asymptotic performance of the
proposed jamming schemes. As the transmitted power $P_S$ increases
to a relatively high value, it yields
\begin{align}
 \lim_{P_S \to \infty}\Gamma_i  = \frac{{L|h_{S_j,R} |^2 |h_{R,S_i} |^2 }}{{|h_{J_1,R} |^2 |h_{R,S_i} |^2  + |h_{S_1,R} |^2 |h_{J_2,S_i} |^2  + |h_{S_2,R} |^2 |h_{J_2,S_i} |^2 }},
 \end{align}
 \begin{align}
 \lim_{P_S \to \infty}\Gamma_{E_i}  = \frac{{|h_{S_i,E} |^2 }}{{|h_{S_j,E} |^2 }} + \frac{{|h_{S_i,R} |^2 }}{{|h_{S_j,R} |^2}}.
 \end{align}

It is clear that both $\Gamma_i$ and $\Gamma_{E_i}$ are independent
of $P_S$, which means that for high $P_S$, the ergodic secrecy rate
$\mathbb{E}\left[ R_S \right]$ stops increasing and converges to a
fixed value. Consider the asymptotic performance of the OS scheme
that grows linearly with the increment of $P_S$ as described by
(\ref{slope_approx}), it is safe to predict that there will be a
crossover point, $P'$, between the ergodic secrecy rate v.s.
transmitted power curve with jamming and the one with non-jamming.
In a power range below $P'$, the jamming scheme outperforms the
non-jamming one, while above this point, the jamming scheme loses
its advantage in providing higher ergodic secrecy capacity.

We note that the analysis above can apply to any scheme with
continuous jamming (i.e., OS-MSISR, OS-MMISR, SS-MSISR, and
SS-MMISR), which indicates that they share the same asymptotic
behavior as the $P_S$ increases. In another word, the proposed
selection techniques (except for OSW and SSW) behave better than the
non-jamming schemes only within a certain transmitted power range.
Fortunately, in a practical case, $P_S$ is always limited in a
relatively low range and will not increase infinitely.

\subsection{Secrecy Performance with Sparsely Distributed Intermediate Nodes}

This is a common configuration in which the the eavesdropper $E$ has
similar distance with two sources $S_1$ and $S_2$ and the
intermediate nodes spread randomly within the considered area. With
a relatively far distance in between, the interference link between
$J_1$ and $R$ becomes weak. As predicted in the previous subsection,
within a certain transmitted power range (less than the crossover
point $P'$), the selection approaches with continuous jamming are
able to provide a higher ergodic secrecy rate than the non-jamming
schemes. This gain proves the introduction of jamming in selection
schemes as an effective technique. Outside this range, the secrecy
rates of the conventional non-jamming schemes continue to grow with
a slope of 0.3322 as verified by (\ref{slope_approx}), whereas those
of the continuous jamming schemes converge to a fixed value. Inside
this scope, the jamming schemes lose their efficiency in providing a
better secrecy performance for the system.

We note that in some particular scenarios, the system's integrated
secrecy performance is not measured by the sum of the total secrecy
rates, but by the minimum secrecy capacity of all the source nodes
in the system. In this situation, OS-MMISR and SS-MMISR can optimize
the overall secrecy performance of the whole system. For the hybrid
schemes, the OSW and SSW schemes are able to provide better secrecy
performance in the whole transmitted power scope, since it overcomes
the bottleneck caused by negative interference on the
relay-destination links.

\subsection{Secrecy Performance With a Close Cluster of the Intermediate Nodes }

Under the condition that all the intermediate nodes are located very
close to each other, we note that the continuous jamming selections
will lose its efficiency in meeting the secrecy constraints.
Specifically, we will discuss two extreme situations in which the
intermediate nodes cluster is near to one of the destination nodes
$S_i$, and to the eavesdropper $E$, respectively.

\subsubsection{The intermediate nodes cluster locates near to one of the destinations}

There are two reasons that make the proposed jamming schemes
inefficient. Firstly, the nodes of the relay/jammer cluster gather
too close to each other, such that the selected jammer in the first
phase $J_1$ has too much negative impact on the selected relay $R$,
which further decreases the SINRs in the second phase. Secondly, the
jamming code from $J_2$ in the second phase also has an
overly-strong interference on the destination to the one it stays
close with.

\subsubsection{The intermediate nodes cluster locates near to the eavesdropper}
Aside from the first reason presented above, in this configuration,
the direct link between the relay $R$ and the eavesdropper $E$ gets
too strong, which will seriously sabotage the secrecy performance of
selection with continuous jamming.

On the other hand, the hybrid protocols (OSW and SSW) will still be
the most effective schemes in this configuration, since the system's
secrecy performance considered here is measured by the ergodic
secrecy rate.

\subsection{Secrecy Performance With the Eavesdropper Near to One of the Source Nodes}

This is the situation in which the eavesdropper $E$ can get the
communicating information most easily, since the direct link between
$E$ and any one of the source nodes is strong, which makes the
introduction of jamming very necessary. The jamming schemes should
be efficient within quite a large power range, and the hybrid
schemes should still perform as the best selection techniques within
the whole power scope.

\section{Numerical Results}

\begin{figure}[ht]
\centering
\includegraphics[width=4.8in]{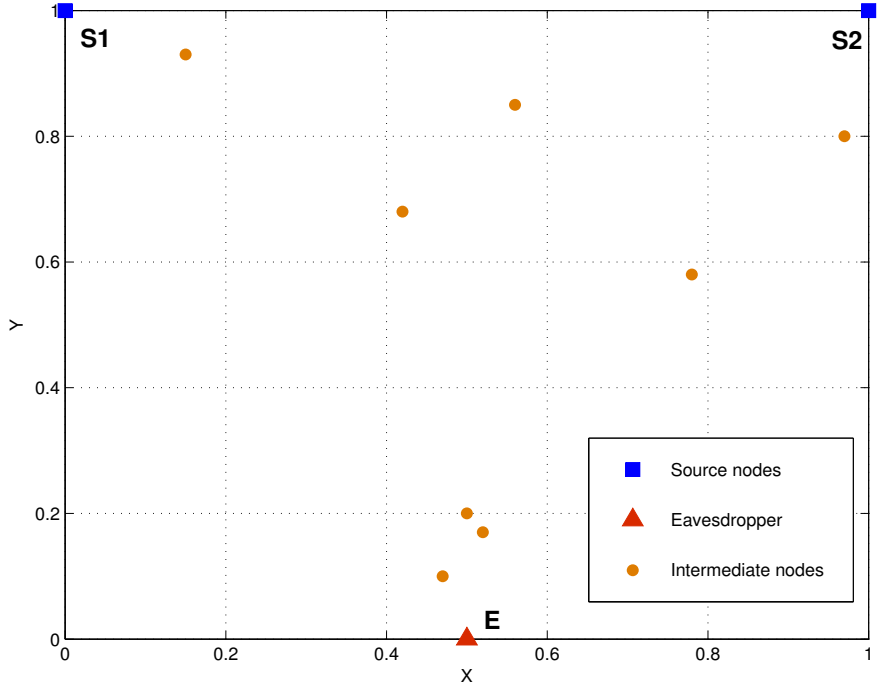}
\caption{The $1\times 1$ simulation environment with $K=8$,
$\beta=3$.} \label{fig_2}
\end{figure}

In this section, we will provide computer simulations in order to
validate the analysis in the previous section. The simulation
environment takes into account two sources $S_1$ and $S_2$, one
eavesdropper $E$, and a intermediate node cluster consisting of
$K=8$ nodes. All the nodes are located in a 2D square topology
within a $1\times 1$ unit square. For simplicity, the source nodes
and the relay transmit with the same power, i.e. $P_S=P_R$. The
relay and jammer nodes transmit with a relay-jammer power ratio
$L=10$. As assumed in Section \Rmnum{2}, the power of the AWGN is
$\sigma^2=1$. The path-loss exponent is set to $\beta=3$. In this
paper, the adopted performance metric is the ergodic secrecy rate.
Meanwhile some results are also provided in terms of secrecy outage
probability $\mathbb{P}\left[ {R_S}\left( {R^*,J_1^* ,J_2^* }
\right)< R_T \right]$, where $\mathbb{P}\left[\cdot \right]$ denotes
probability, and $R_T$ is the target secrecy rate.

In the first simulation, we assume a scenario where $S_1$, $S_2$ and
$E$ are located at $\left( {X_E ,Y_E } \right) = \left( {0.5,0}
\right)$, $ \left( {X_{S_1} ,Y_{S_1} } \right) = \left( {0,1}
\right)$ and $ \left( {X_{S_2} ,Y_{S_2} } \right) = \left( {1,1}
\right)$, respectively. The intermediate nodes spread randomly
within the square space, as shown in Fig.~\ref{fig_2}.

\begin{figure}[ht]
\centering
\includegraphics[width=4.8in]{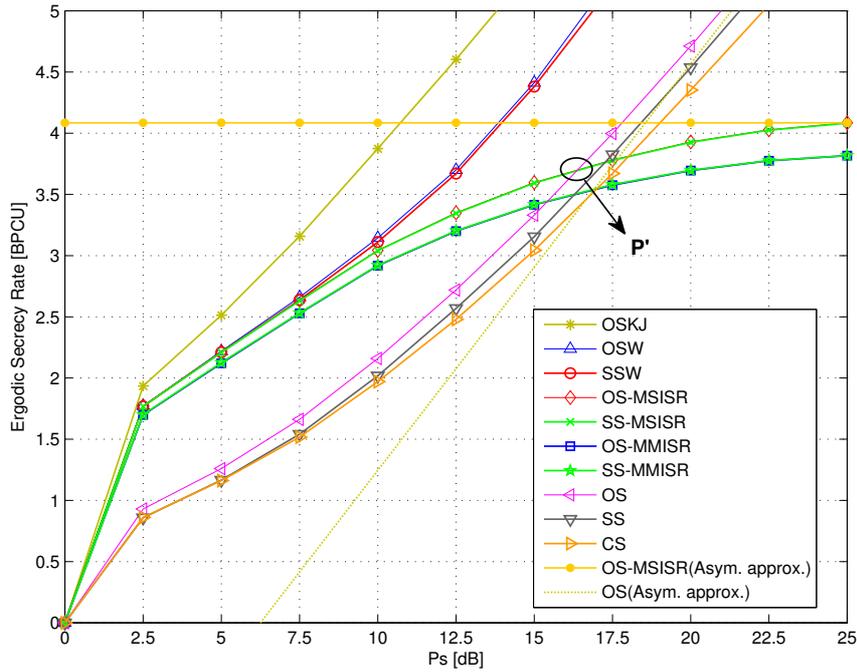}
\caption{Ergodic secrecy rate versus transmitted power $P_S$ for
different selection techniques.} \label{fig_3}
\end{figure}

Fig.~\ref{fig_3} shows the ergodic secrecy rate versus the
transmitted power $P_S=P_R$ of different selection schemes. We can
observe that selection algorithms with jamming outperform their
non-jamming counterparts within a certain transmitted power range
(less than $P'\approx 16dB$), where the ergodic secrecy rate of the
OS-MSISR scheme is approximately higher than that of the OS scheme
by 1 bit per channel use (BPCU).  Outside this range ($P>P'$), the
secrecy rate of OS-MSISR converges to a power-independent value
which is approximately 4.1 BPCU, whereas the ergodic secrecy rate of
OS continues to grow with a slope of 0.3322 , as proved by
(\ref{slope_approx}). This validates the secrecy performance
analysis in Section \Rmnum{4}. In addition, we can see that in this
relay topology, the suboptimal schemes (SS-MSISR, SS-MMISR) which
are based on average channel knowledge perform almost the same as
the optimal schemes (OS-MSISR, OS-MMISR), which implies that in this
configuration where the intermediate nodes are sparsely distributed,
an average channel knowledge may also provide enough information in
order to get optimal relay selection.

\begin{figure}[ht]
\centering
\includegraphics[width=4.8in]{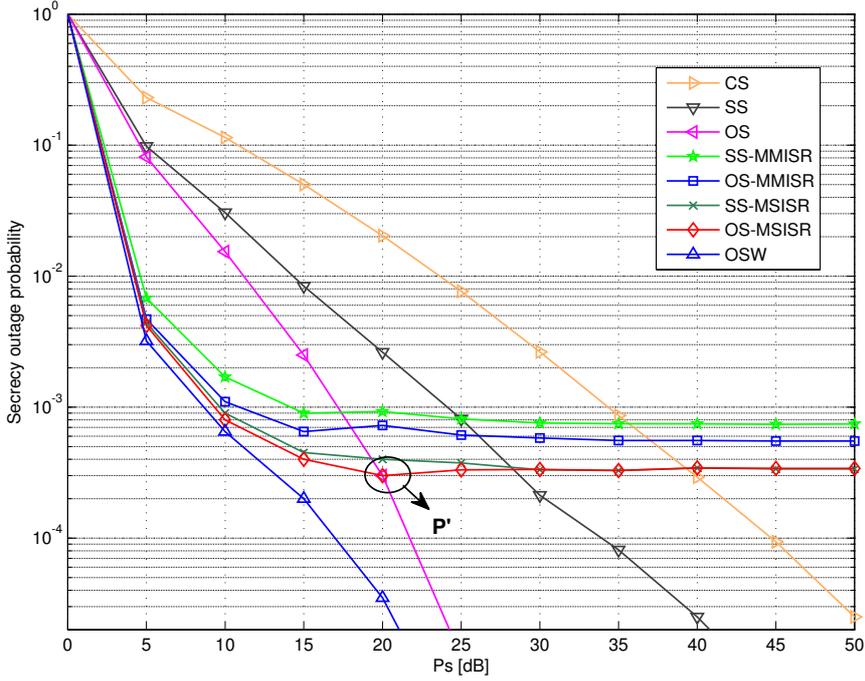}
\caption{Secrecy outage probability versus transmitted power $P_S$
for different selection techniques with $R_T=0.2$ BPCU.}
\label{fig_4}
\end{figure}

In Fig.~\ref{fig_3}, a comparison between the OS-MSISR and OS-MMISR
shows that the OS-MSISR scheme has slightly higher ergodic secrecy
capacity by about 0.25 BPCU than OS-MMISR does corresponding to
transmitted power $P_S$. The same comparison result can be observed
from the SS-MSISR and SS-MMISR schemes, which matches our previous
analysis. Furthermore, it can be seen that OSW performs better than
any other selection techniques with or without continuous jamming.
At a low power range where $P_S<P'$, the OSW scheme performs
slightly better than OS-MSISR, but much better than OS (by about 1.2
BPCU), for the reason that in this range continuous jamming is
almost always needed. After $P_S$ grows much higher than $P'$, OSW
outperforms both the other two schemes by a large scale. For the
suboptimal case, we can see that SSW provides almost the same
performance as the OSW scheme in this relay topology, which
validates the practical value of this hybrid scheme. An observation
of the performance of OSKJ scheme shows that it outperforms all the
other selection techniques, providing the highest ergodic secrecy
rate when the transmitted power increases due to its ability of the
destinations to decode the artificial interference in this OSKJ
scheme.

\begin{figure}[ht]
\centering
\includegraphics[width=4.8in]{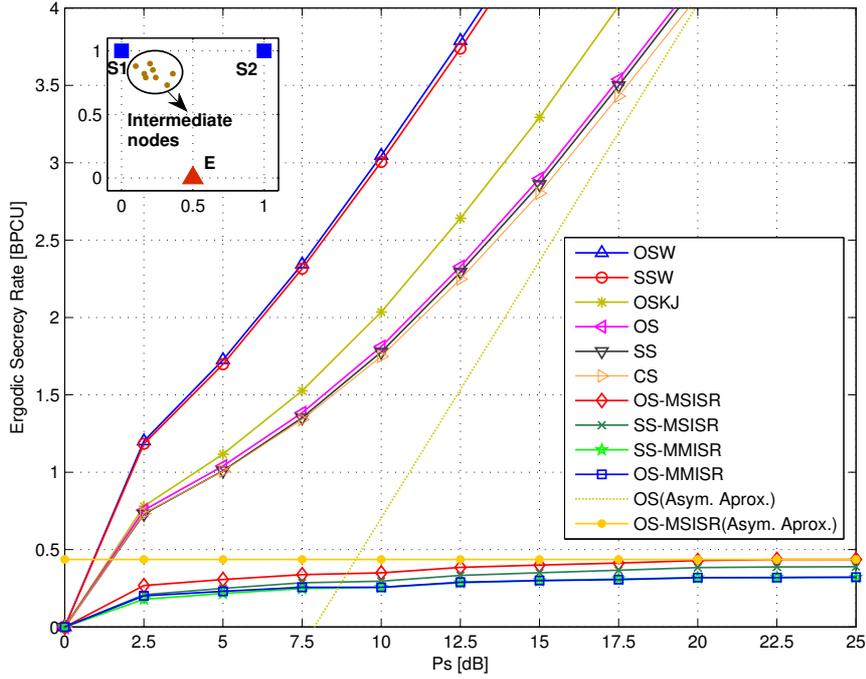}
\caption{Ergodic secrecy rate for a scenario where the intermediate
nodes are close to $S_1$.} \label{fig_5}
\end{figure}

Within this configuration, we also compare the performance of
different selection techniques measured by secrecy outage
probability, which is shown in Fig.~\ref{fig_4}. The target secrecy
rate $R_S$ is set as 0.2 BPCU. It can be seen that selection schemes
with jamming provies lower secrecy outage probability within a
certain transmitted power range ($P_S<P'$, $P'\approx 20 dB$).
Outside this range, the conventional selection without jamming
achieves better secrecy outage probability. Regarding the hybrid
protocols, the OSW scheme outperforms the non-switching selection
techniques.

In Fig.~\ref{fig_5}, it deals with a configuration where the
intermediate nodes cluster, which also includes $K=8$ nodes, is
located closely near to one of the two users (e.g., node $S_1$,
without loss of generality). We can see the ergodic secrecy rate of
the proposed selection schemes in this topology differs greatly from
that in the previous configuration. We observe that continuous
jamming schemes (i.e. OS-MSISR, OS-MMISR, SS-MSISR, and SS-MMISR)
are inefficient here, which converge to less than 0.5 BPCU,
validating our discussion in Section \Rmnum{4}.

\begin{figure}[ht]
\centering
\includegraphics[width=4.8in]{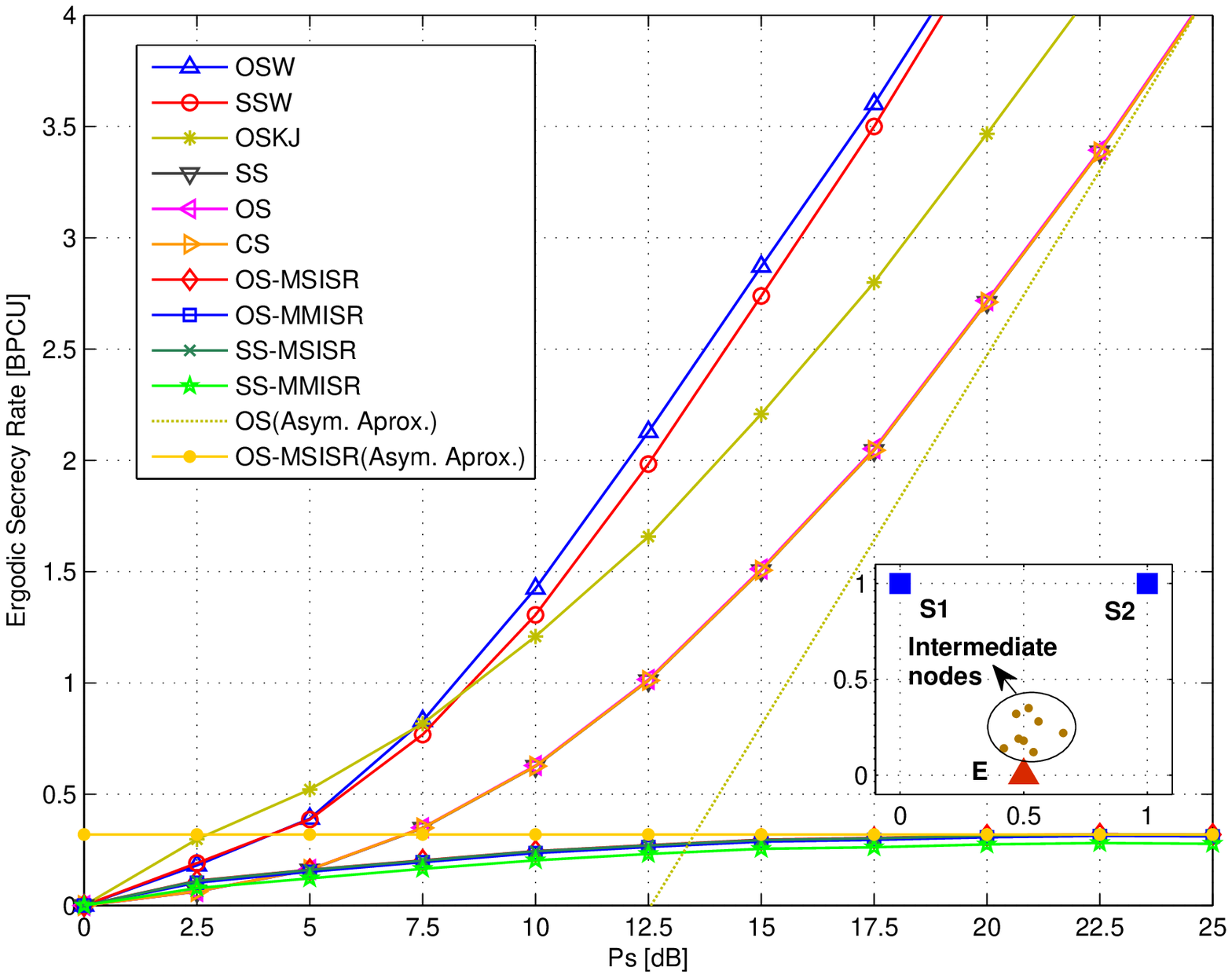}
\caption{Ergodic secrecy rate for a scenario where the intermediate
nodes are close to the eavesdropper $E$.} \label{fig_6}
\end{figure}

On the other hand, OSW and SSW still outperform all the other
selection techniques by a quite large scale (more than 4 BPCU when
$P_S$ is very high, as shown in Fig.~\ref{fig_5}. We also note that
in this topology, the OSW and SSW schemes perform even better than
the OSKJ scheme, which seems to be an interesting result. Further
investigation reveals that the involvement of $J_1$ node in OSKJ
causes a different value of $\alpha$ with that of OSW and SSW, which
results in lower secrecy rates in OSKJ than in OSW and SSW schemes.
This indicates that the proposed OSW/SSW schemes may perform even
better than the ``ideal'' case where the destinations can mitigate
the artificial interference. All of these validate the value of the
selection techniques with intelligent switching in potential
practical use.

In Fig.~\ref{fig_6}, we set the intermediate nodes cluster closely
to the eavesdropper $E$. Here the jamming schemes also perform worse
than non-jamming ones in most of the transmitted power range. It
also shows the range where continuous jamming schemes perform better
than non-jamming schemes in this topology is slightly larger than
that of the previous one, since there is no strong $R\to E$ link
here. Regarding to the hybrid schemes, OSW and SSW still perform as
the best selection techniques in providing the highest secrecy rate.

\begin{figure}[ht]
\centering
\includegraphics[width=4.8in]{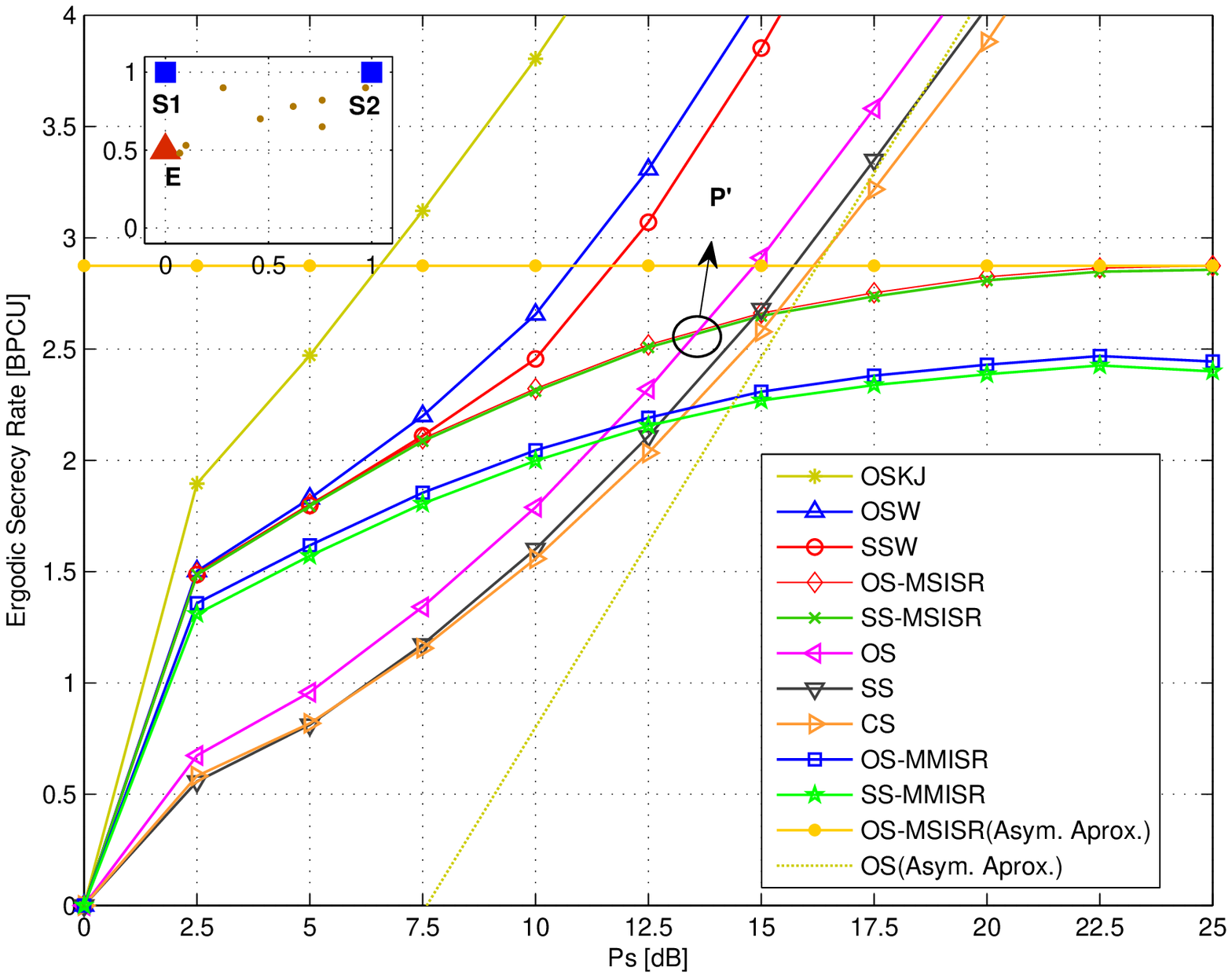}
\caption{Ergodic secrecy rate for a scenario where the eavesdropper
$E$ is close to $S_1$.} \label{fig_7}
\end{figure}

Finally, we place the eavesdropper $E$ near to one of the two
sources (taken $S_1$ for example) to examine the results. The
location of eavesdropper $E$ is set to $ \left( {X_E ,Y_E } \right)
= \left( {0,0.5} \right)$, the intermediate nodes are spread
randomly across the considered rectangle area, as shown in the inset
of Fig.~\ref{fig_7}. We get a similar simulation result with that of
the first configuration, in which the eavesdropper $E$ has the same
distance with $S_1$ and $S_2$. The non-jamming schemes (CS, OS and
SS) here are less effective in promoting the secrecy performance. On
the contrary, the selection techniques with continuous jamming
(OS-MSISR, OS-MMISR, SS-MSISR and SS-MMISR) provide much higher
secrecy capacity in a large transmitted power range ($P' \approx 13
dB$). Within this power range, the hybrid schemes (OSW and SSW)
perform slightly better than the continuous jamming techniques
because jamming is almost always needed in this configuration.
Outside this regime, where the non-jamming scheme performs better,
the difference between the intelligent switching and continuous
jamming increases and the hybrid schemes still perform as the most
efficient schemes.

\section{Conclusions}

This paper has studied the joint relay and jammer selection in
two-way cooperative networks with physical layer secrecy
consideration. The proposed scheme achieves an opportunistic
selection of one conventional relay node and one (or two) jamming
nodes to increase security against eavesdroppers based on both
instantaneous and average knowledge of the eavesdropper channels.
The selected relay node helps enhance the information transmission
between the two sources via an AF strategy, while the jamming nodes
are used to produce intentional interference at the eavesdropper
nodes in different transmission phases. We found that the proposed
jamming schemes (i.e. OS-MSISR, OS-MMISR, SS-MSISR, and SS-MMISR)
are effective within a certain transmitted power range for scenarios
with sparsely distributed intermediate nodes. Meanwhile the
non-jamming schemes (CS, OS, and SS) are preferred in configurations
where the intermediate nodes are confined close to each other. The
OSW scheme which switches intelligently between jamming and
non-jamming modes is very efficient in providing the highest secrecy
rate in almost the whole transmitted power regime in two-way
cooperative networks, but it requires an instantaneous eavesdropper
channel knowledge. On the other hand, the suboptimal switching
scheme, SSW, which is based on the average knowledge of the
eavesdropper channel and therefore much practical, provides a
comparable secrecy performance with the OSW scheme.

\ifCLASSOPTIONcaptionsoff
  \newpage
\fi

\pagebreak

\end{document}